\documentclass[11pt]{article}
\usepackage{graphics,amssymb}
\usepackage{epsfig}
 
\begin{document}
 
{
\setlength{\textwidth}{16.5cm}
\setlength{\textheight}{22.2cm}
\setlength{\hoffset}{-1.43cm}
\setlength{\voffset}{-.9in}

%%%%% The following lines create the SLAC Pub Title Page
%%
\thispagestyle{empty}
\renewcommand{\thefootnote}{\fnsymbol{footnote}}

%%%%% Substitute your Pub number, month and year in the following:
%%
\begin{flushright}
{\normalsize
SLAC-AP-135\\
ATF-00-14\\
December 2000}
\end{flushright}

\vspace{.8cm}

%%%%% Title and Author Information:
%%
\begin{center}
{\bf\Large Intrabeam Scattering Analysis of\break  
ATF Beam Data Taken in April 2000
\footnote{\small Work supported by
Department of Energy contract  DE--AC03--76SF00515.}}

\vspace{1cm}

{\large
K.L.F. Bane\\
Stanford Linear Accelerator Center, Stanford University,
Stanford, CA  94309}

\medskip

{\large
H. Hayano, K. Kubo, T. Naito, T. Okugi, and J. Urakawa\\
High Energy Accelerator Research Organization (KEK), Tsukuba, Japan}

\end{center}
%\vfill 

\begin{center}
{\bf\large   
Abstract }
\end{center}

\begin{quote}

Our theoretical comparisons
suggest that the ATF measurement results of April 2000 for 
energy spread, bunch length, and horizontal emittance {\it vs.}
current, 
 and a low current emittance ratio of about
1\% are generally consistent with intrabeam scattering (IBS) theory,
though the measured effects appear to be 
stronger than theory. In particular, 
the measured factor of 3
growth in vertical emittance at 3~mA
does not seem to be supported.
It appears that 
either (1)~there is another, unknown force in addition to IBS causing
emittance growth in the ATF;
or (2)~the factor of 3 vertical emittance growth is not real, and
our other discrepancies
are due to the approximate nature of IBS theory.
Our results suggest, in addition, that, even if IBS theory
has inaccuracies, the effect will be
useful as a diagnostic in the ATF.
For example, by measuring the energy spread, one
will be able to obtain the emittances.
Before this can be done, however, 
more benchmarking measurements will be needed.

\end{quote}

}
\vfill

\title{Intrabeam Scattering Analysis of  
ATF Beam Data Taken in April 2000}
\author{K.L.F. Bane, H. Hayano, K. Kubo, T. Naito, T. Okugi, and J. Urakawa}
\date{}
\maketitle

\section*{Introduction}

Between April 13-19, 2000 single bunch energy spread, bunch length, and 
horizontal and
vertical emittances in the ATF were all measured
as a function of current\cite{april}-\cite{emittanceb}. One surprising
outcome was that the vertical emittance appeared to grow by
a factor of 3 by a current of 3~mA. 
The ATF is a prototype damping ring for the JLC/NLC linear colliders,
and the concern with this result is that  
%if there indeed is an unexpectedly large growth in vertical emittance
%in such storage rings, the effect on linear collider performance
it may portend
an as yet not understood and unexpected growth in such damping rings,
which would have negative ramifications on collider performance.
However, since the $x$-$y$ coupling
in the ATF 
is very small ($\sim1\%$), and since the emittance measurements
were performed on the beam after it had been extracted from the ring,
the question was, How much of this measured $y$-emittance growth
was real and how much was due to measurement error, such as 
dispersion in the extraction line or in the wire monitors
used for the measurements. 

With the ATF as it is now, running below design energy
and with the wigglers turned off, the beam properties are strongly affected
by intra-beam scattering (IBS), an effect that couples the
three dimensions of the beam together. 
In April 2000 all the beam dimensions were measured
to varying degrees of accuracy, and
the hope is that the knowledge
of IBS theory can be used to check for consistency in
the data. 
Besides the question of the vertical emittance growth,
we hope that IBS theory can be used in beam diagnostics in the future.
For example, the energy spread measurement is quick, accurate, and
easy to do. It would be nice if this measurement could be used to
estimate the beam emittance and/or bunch length, beam properties that
are much more difficult to measure directly. 

The literature on intrabeam scattering is quite extensive.
(For an introduction, see for example the IBS section and its bibliography,
written by A.~Piwinski, in the
{\it Handbook of Accelerator Physics and Engineering}
~\cite{handbook}).
The first rather thorough treatment of IBS in circular accelerators is
due to Piwinski (P), derived following a 
two-particle Coulomb scattering formalism\cite{Piw}. Another formalism is
that of Bjorken-Mtingwa (B-M), obtained following
quantum mechanical two particle scattering rules\cite{BM}. 
This is the formalism that is more often used in modern optics
programs that also calculate IBS, programs such as SAD\cite{SAD}. 
The B-M result is considered to be more general in that
the combination of optics terms 
$\beta_x\phi_x\equiv(\beta_x\eta_x^\prime-\beta_x^\prime\eta_x/2)$
around the ring
do not need to
be small compared to $\eta_x$, whereas in the P 
method it seems they do\cite{Piwref}.
(Note that this condition is typically violated in
modern low emittance storage rings.)
The B-M formalism, however, does not include vertical dispersion,
whereas the P formalism does.   
Neither formalism includes $x$-$y$ coupling, 
though a more generalized formulation, which includes both
linear coupling and can also be applied to low emittance machines,
is given by Piwinski in Ref.~\cite{Piw2}.
Note that in deriving such formulations many approximations were made,
having to do with the cut-off distance for scattering
(typically taken to be the vertical beam size), 
curved trajectories, etc. 
In addition, it appears
that no current IBS formalism properly
accounts for the effects of potential well
distortion or of the micro-wave instability.
Finally, T.~Raubenheimer has pointed out that IBS does not result
in Gaussian bunch distributions, and that these theories give rms beam
sizes that can greatly overemphasize few particles occupying 
the tails\cite{Tor}.
% though a theoretical
%study on the subject can be found\cite{IBSPW}.

Most of the early papers comparing IBS theory and measurement
were for bunched and unbunched hadronic machines, where the effect
tends to be more pronounced than in high energy electron machines.
For example, Conte and Martini, for unbunched protrons
in the CERN Antiprotron Accumulator ring, 
found good agreement for the longitudinal and horizontal IBS 
growth rates, but no agreement for the vertical rate\cite{Conte}. 
Evans and Gareyte, for bunched protrons in the SPS, found good agreement
for the radial emittance growth rate with time, once an additional factor
representing gas scattering was included\cite{Evans}.
As for electron machines, both IBS and the related Touschek effect have become
important effects in modern, low emittance light sources. 
C.H.~Kim at the ALS
found that he can get agreement with measured horizontal emittance
growth with current, but to do this he needed to include a 
significant additional fitting factor in the calculations\cite{CKim}. 
So it may be too much to expect good agreement between IBS
theory and measurement without the use of such fudge factors.
Indeed, A.~Piwinski has said that, given the approximations
taken in deriving IBS theory, one can expect agreement between
theory and measurement only on the order of a factor of 2\cite{Piwref}.
Yet even if this became the case, it may still be possible to 
benchmark the ATF with accurate emittance and energy spread measurements,
to find the fudge factors. Then in the future, 
one may be able to perform simpler measurements---like the energy
spread measurement---to get an estimate
of parameters that are more difficult to obtain directly,
such as bunch length and emittance.

\section*{Piwinski's Solution}

We are interested in what happens at low coupling, and we
will concentrate on using the more simplified version of
Piwinski's solution. 
We begin by reproducing the Piwinski solution in its 
entirety\cite{handbook}. Note that there is nothing new in this
section, except the way potential well distortion is
added to the calculation. 

Let us consider the IBS growth rates 
in energy $p$, 
in the horizontal direction $x$, and in the vertical direction 
$y$ to be defined as 
\begin{equation}
{1\over T_p}={1\over\sigma_p}{d\sigma_p\over dt}\ ,\quad
{1\over T_x}={1\over\epsilon_x^{1/2}}{d\epsilon_x^{1/2}\over dt}\ ,\quad
{1\over T_y}={1\over\epsilon_y^{1/2}}{d\epsilon_y^{1/2}\over dt}\ .
\end{equation} 
Here $\sigma_p$ is the rms (relative) energy spread, 
 $\epsilon_x$ the horizontal emittance, and
$\epsilon_y$ the vertical emittance.
According to Piwinski the IBS growth rates are given as
\begin{equation}
{1\over T_p}=\left< A{\sigma_h^2\over\sigma_p^2}f(a,b,q)\right>
\label{tp}
\end{equation} 
\begin{equation}
{1\over T_x}=\left< A\left[f({1\over a},{b\over a},{q\over a})
+{D_x^2\sigma_h^2\over\sigma_{x\beta}^2}f(a,b,q)\right]\right>
\label{tx}
\end{equation} 
\begin{equation}
{1\over T_y}= \left< A\left[f({1\over b},{a\over b},{q\over b})
+{D_y^2\sigma_h^2\over\sigma_{y\beta}^2}f(a,b,q)\right]\right>
\label{ty}
\end{equation}
where the brackets $\langle\rangle$ mean that the enclosed
quantities, combinations of beam parameters and lattice
properties,
 are averaged around the entire ring. Parameters are: 
\begin{equation}
A= {r_0^2 N\over 64\pi^2\beta^3\gamma^4\epsilon_x\epsilon_y\sigma_s\sigma_p}
\end{equation} 
\begin{equation}
{1\over\sigma_h^2}= {1\over\sigma_p^2} + 
{D_x^2\sigma_h^2\over\sigma_{x\beta}^2} +
{D_y^2\sigma_h^2\over\sigma_{y\beta}^2}
\end{equation} 
\begin{equation}
a={\sigma_h\beta_x\over\gamma\sigma_{x\beta}},\quad
b={\sigma_h\beta_y\over\gamma\sigma_{y\beta}},\quad
q=\sigma_h\beta\sqrt{{2d\over r_0}};
\end{equation}
The function $f$ is given by:
\begin{equation}
f(a,b,q)=8\pi\int_0^1\left\{2\ln\left[{q\over2}\left({1\over P}+
{1\over Q}\right)\right] -0.577\ldots\right\}{1-3u^2\over PQ}\,du
\end{equation} 
\begin{equation}
P^2= a^2+(1-a^2)u^2,\quad\quad Q^2= b^2+(1-b^2)u^2
\end{equation} 

The global beam properties 
that are affected by IBS are the rms (relative) energy spread
$\sigma_p$ and the rms bunch length 
$\sigma_s$, the horizontal emittance $\epsilon_x$, and
the vertical emittance $\epsilon_y$. Other global properties
are the bunch population $N$, the relative velocity $\beta$, and
the energy factor $\gamma$.
 Note that $r_0$ is the classical radius of the
particles (for electrons $r_0=2.82\times10^{-15}$~m). The lattice
functions needed are the beta functions $\beta_x$, $\beta_y$, and
the dispersion functions $D_x$, $D_y$. Note also that
$\sigma_{x\beta}=\sqrt{\beta_x\epsilon_x}$, and 
$\sigma_{y\beta}=\sqrt{\beta_y\epsilon_y}$.
The parameter $d$ represents a cut-off for
the IBS force, which Piwinski says should be taken as the vertical
beam size, but he also points out that the results are not
very sensitive to exactly what is chosen for this parameter.

%\begin{equation}
%\sigma^2_{x\beta,y\beta,p}={\sigma^2_{x\beta0,y\beta0,p0}\over
%1-\tau_{x,y,p}/T_{x,y,p}}
%\end{equation} 

Then the steady-state beam properties are given by
\begin{equation}
\epsilon_x={\epsilon_{x0}\over 1-\tau_x/T_x}\ ,\
\epsilon_y={\epsilon_{y0}\over 1-\tau_y/T_y}\ ,\
\sigma^2_{p}={\sigma^2_{p0}\over
1-\tau_{p}/T_{p}}\ ,
\label{eqiterate}
\end{equation} 
where subscript 0 represents the beam property due to synchrotron
radiation alone, {\it i.e.} in the absence of IBS, and $\tau_x$,
$\tau_y$, and $\tau_p$ are the synchrotron radiation damping times
in the three directions. These are 3 coupled equations
in that all 3 IBS rise times depend on $\epsilon_x$,
$\epsilon_y$, and $\sigma_p$. Note 
that a 4th equation,
the relation between $\sigma_s$ and $\sigma_p$, is also implied;
generally this is taken to be the nominal (zero current)
relationship.

%\begin{equation}
%{d\sigma^2_{x\beta,y\beta,p}\over dt}= -{1\over\tau_{x,y,p}}
%\left(\sigma^2_{x\beta,y\beta,p}-\sigma^2_{x\beta0,y\beta0,p0}\right)
%+{1\over T_{x,y,p}}\sigma^2_{x\beta,y\beta,p}
%\end{equation} 

%\begin{equation}
%{d\sigma^2_{x\beta,y\beta,p}\over dt}=
%-{(\sigma^2_{x\beta,y\beta,p}-\sigma^2_{x\beta0,y\beta0,p0})\over
%\tau_{x,y,p}}
%+{\sigma^2_{x\beta,y\beta,p}\over T_{x,y,p}}
%\end{equation} 

Piwinski suggests iterating Eqs.~\ref{eqiterate}
 until a self-consistent
solution is found. Our experience is that this has the problem that
negative values of emittance or $\sigma_p^2$ can be obtained
with this procedure, causing
difficulty in knowing how to 
continue the iteration. We find that a better method
is to convert these equations into 3 coupled differential equations,
such as is done in Ref.~\cite{CKim} to obtain the time development of the
beam properties in a ring. 
Here, however, we use it only as a mathematical
device for
 finding the steady-state solution. For example, the equation
for $\epsilon_x$ becomes 
\begin{equation}
{d\epsilon_x\over dt}=
-{(\epsilon_x-\epsilon_{x0})\over
\tau_x}
+{\epsilon_x\over T_x}\quad,
\end{equation} 
and there are corresponding
 equations for $\epsilon_y$ and $\sigma_p^2$.

We have three comments:
(1)~In a storage ring 
vertical emittance is generally the result of two phenomena, $x$-$y$ 
coupling---caused, for example, 
by rolled magnets---and vertical dispersion---caused 
by vertical closed orbit distortion. 
Although our formalism technically is
valid only for the case of no $x$-$y$ coupling, we believe
that it can also be used for the case of weak coupling. In such a
case we simply pick a zero-current emittance ratio
$r_{xy0}=\epsilon_{y0}/\epsilon_{x0}$, and then set
$\epsilon_{y0}$ accordingly. Note that $r_{xy0}$ is meant to include
both the contributions of $x$-$y$ coupling and of vertical dispersion.
Finally, note that the vertical emittance, assuming only vertical
dispersion, can be approximated by\cite{Torth}
\begin{equation}
\epsilon_{y0}= 2{\cal J}_\epsilon{(D_y)_{rms}^2\over
\langle\beta_y\rangle}\sigma_{p0}^2\quad,
\label{epsy0}
\end{equation}
with ${\cal J}_\epsilon$ the energy damping partition number.

(2)~It is not clear how impedance effects and IBS effects interact.
At the ATF we have shown that up to the 
currently highest attainable single bunch currents ($\sim3$~mA)
a micro-wave threshold has not yet been reached\cite{april}.
But we do appear to have a sizeable potential well bunch 
lengthening effect. We will 
assume that this case can be approximated by adding the
proper multiplicative
factor $f_{pw}(N)$, obtained from measurements,
 to the equation giving $\sigma_s$ in terms of
$\sigma_p$. 
(3)~As mentioned earlier, the IBS bunch distributions are not
Gaussian, and tail particles can be overemphasized in the 
beam size solutions
given above. T.~Raubenheimer in Ref.~\cite{Tor} gives a way of estimating
IBS beam sizes that better reflect the 
particles in the core of the beam.
Note that the optics computer program SAD can solve the
IBS equations of B-M, include the effects of orbit errors,
and also find the sizes of the core of the beam.

\section*{Parameter Studies}

We have programmed the above Piwinski equations.
We have also programmed the B-M equations, which also
give 3 IBS growth rates (though there is a
factor of 2 difference in their definition). 
Let us begin our numerical
studies by comparing the results of the two programs
when applied to the ATF lattice and beam properties.
(Note that there is a published comparison of 
results of these two methods applied to the CERN AA ring
in Ref.~\cite{Martini}; however, the variation of the lattice
parameters around the ring was not included in the P method
calculation.)
As parameters we take: $E=1.28$~GeV, $\sigma_{p0}=5.44\times10^{-4}$,
$\sigma_{s0}=5.06$~mm (for an rf voltage of 300~kV), 
$\epsilon_{x0}=1.05$~nm, $\tau_p=20.9$~ms, $\tau_x=18.2$~ms,  
and $\tau_x=29.2$~ms. The ATF circumference is 138~m,
$\langle\beta_x\rangle=4.2$~m, $\langle\beta_y\rangle=4.6$~m,
$\langle D_x\rangle=5$~cm. The function $\beta_x|\phi_x|$ is
about $.5D_x$ at positions of the 
minima in $D_x$, and about $.15D_x$ at positions of the maxima,
which are not so small, and we might expect some disagreement
between the two methods.
Note that for the averages represented by brackets 
in Eqs.~\ref{tp}-\ref{ty}, and their counterparts in the B-M
method,
we calculate, as is
normally done, the appropriate combination of lattice and beam
properties at the ends of the lattice elements, connect these
with straight lines, and then find the average of the resulting curve.

\begin{figure}[p]
\centering
\epsfig{file=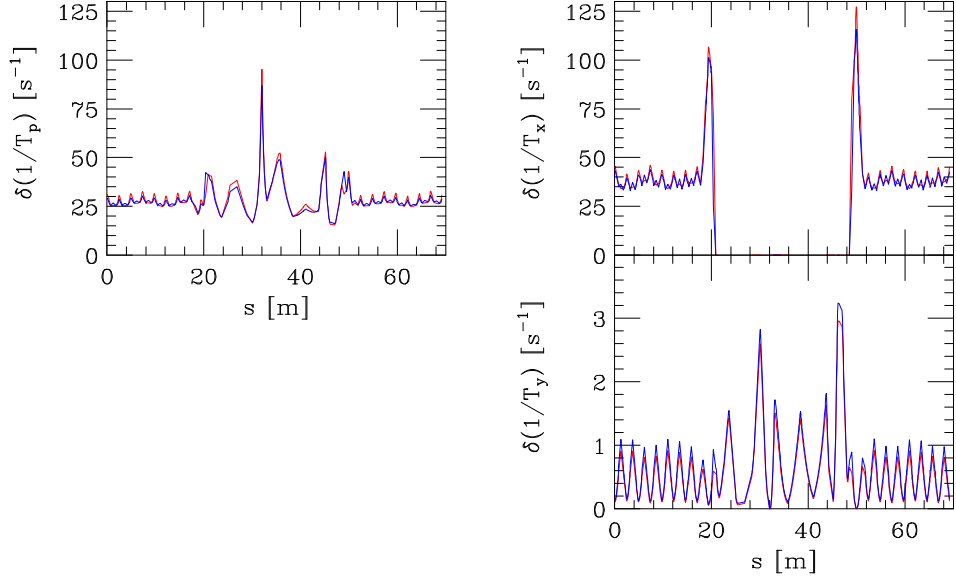, width=12.6cm}
\caption{
Differential growth rates over
1/2 the ATF, as obtained by 
the Piwinski (blue) and the Bjorken-Mtingwa (red) methods 
Here $I=3.1$~mA, $r_{xy0}=.01$, and $D_y=0$ (no vertical dispersion). 
}
\label{fidebug}
\end{figure}
 
\begin{figure}[p]
\centering
\epsfig{file=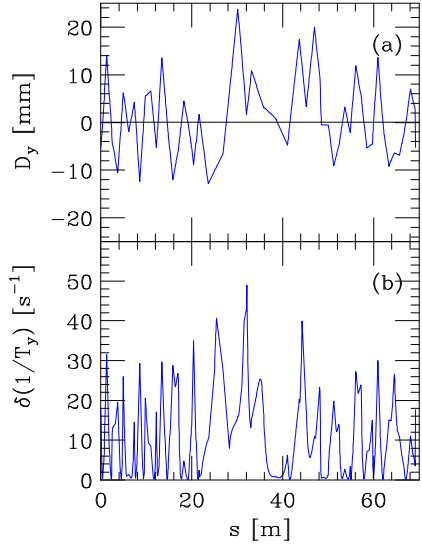, width=5.5cm}%6.5
\caption{
Random vertical dispersion [$(D_y)_{rms}=7.5$~mm]~(a), and the resulting
differential vertical growth rate~(b) over
1/2 the ATF. $1/T_y=11.8$~s$^{-1}$.
Here $I=3.1$~mA and $r_{xy0}=.01$. 
}
\label{fidebugb}
\end{figure}

Fig.~\ref{fidebug} displays the 3 differential IBS growth
rates as obtained by the two methods (blue for P, red for B-M)
 when
$I=3.1$~mA and $r_{xy0}=.01$. 
The IBS
growth rates are just the average values over the ring
of these functions.
Since the ATF has two-fold
symmetry we give the result for only half the ring
(the straight section is in the middle). 
Here we include no vertical dispersion, 
{\it i.e.} the vertical emittance is presumed to be
only due to (weak) $x$-$y$ coupling, and
$f_{pw}$ is set to 1. 
We see almost perfect agreement for the differential
growth rates obtained by the two methods, with only slight
differences in the peaks.
As for the averages, for the Piwinski method $1/T_p=28.4$~s$^{-1}$,
$1/T_x=24.8$~s$^{-1}$, and $1/T_y=.67$~s$^{-1}$. The B-M
results are 28.9, 26.2, and .58~s$^{-1}$, respectively.
Note that the $y$ growth rates are very small.
As for the steady state beam properties, 
for this example $\sigma_p/\sigma_{p0}=1.566$,
$\epsilon_x/\epsilon_{x0}=1.825$, and $\epsilon_y/\epsilon_{x0}=.0102$
for P, and 1.587, 1.910, .0102 for B-M, respectively. We see that
for the ATF
the results of the two methods are almost the same.
Finally, note that when making the same comparison for the ALS at
Berkeley, a low emittance light source with $\beta_x|\phi_x|$
comparable to $D_x$, we again find very close agreement in
the results of the two methods.

In the ATF the rms vertical dispersion, after correction, 
is typically 3-4~mm.
To simulate this effect we added a randomly
generated component of $D_y$ (weighted by $\sqrt{\beta_y}$)
at the high and low $\beta_y$ points, and connected these with
straight lines (see Fig.~\ref{fidebugb}a). This was added 
to the P calculation. 
In the ATF typically $r_{xy0}\approx.01$
and ${\cal J}_\epsilon=1.4$.
We see from Eq.~\ref{epsy0} that if $(D_y)_{rms}=4$~mm,
then 30\% of the low current emittance is due to dispersion,
and the rest due to $x$-$y$ coupling.
Let us consider an example now where 
the size of $\epsilon_{y0}$ is given entirely
by vertical dispersion, with $r_{xy0}=.01$ and
$(D_y)_{rms}=7.5$~mm.
In Fig.~\ref{fidebugb}b we plot
the resulting $\delta(1/T_y)$ at $I=3.1$~mA.
 The growth is much larger than before, with
$1/T_y=11.8$~s$^{-1}$.
(Note that $\sigma_p/\sigma_{p0}=1.48$,
$\epsilon_x/\epsilon_{x0}=1.71$, and $\epsilon_y/\epsilon_{x0}=.015$.)
 We can understand the change in $1/T_y$ if we look
back at the equations for the growth rates in $x$ and $y$,
Eqs.~\ref{tx} and \ref{ty}. In Eq.~\ref{tx} the horizontal growth rate is
dominated by the second term in the brackets, the one proportional
to $D_x^2/\sigma_{x\beta}^2$. In Eq.~\ref{ty} the vertical growth rate,
when $D_y=0$,
is given by the small first term in the brackets. 
Since the second term is proportional to  $D_y^2/\sigma_{y\beta}^2$,
$1/T_y$ will become comparable to $1/T_x$ when 
$(D_y)_{rms}\sim \langle D_x\rangle\sqrt{\epsilon_y/\epsilon_x}$
(since $\langle\beta_x\rangle$ and $\langle\beta_y\rangle$ are similar),
which equals 5~mm. Note that the importance grows as the second
power of $(D_y)_{rms}$.

\begin{figure}[p]
\centering
\epsfig{file=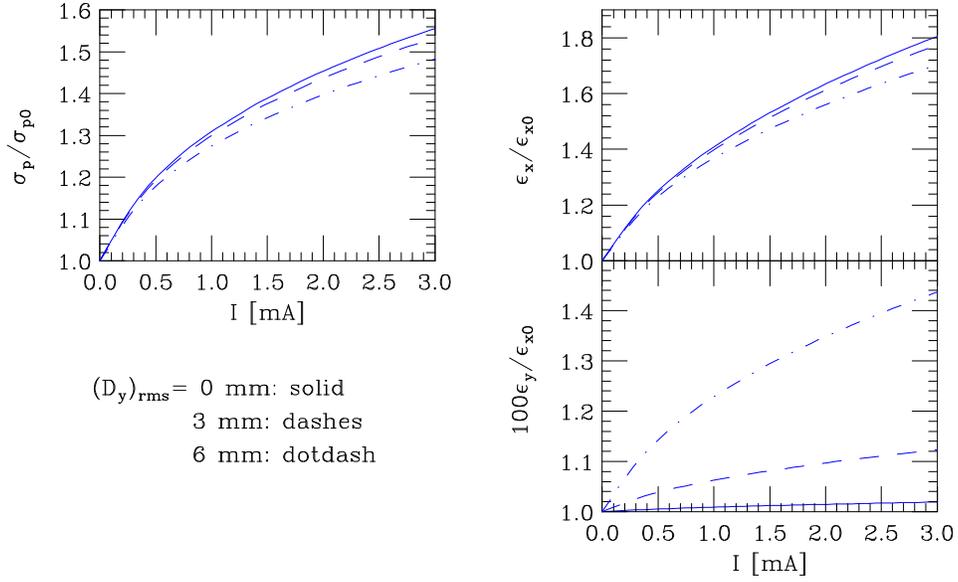, width=12.6cm}
\caption{
Calculations of the current dependence of the beam
properties when $r_{xy0}=.01$, $f_{pw}=1$.
}
\label{fidnv}
\end{figure}

\begin{figure}[p]
\centering
\epsfig{file=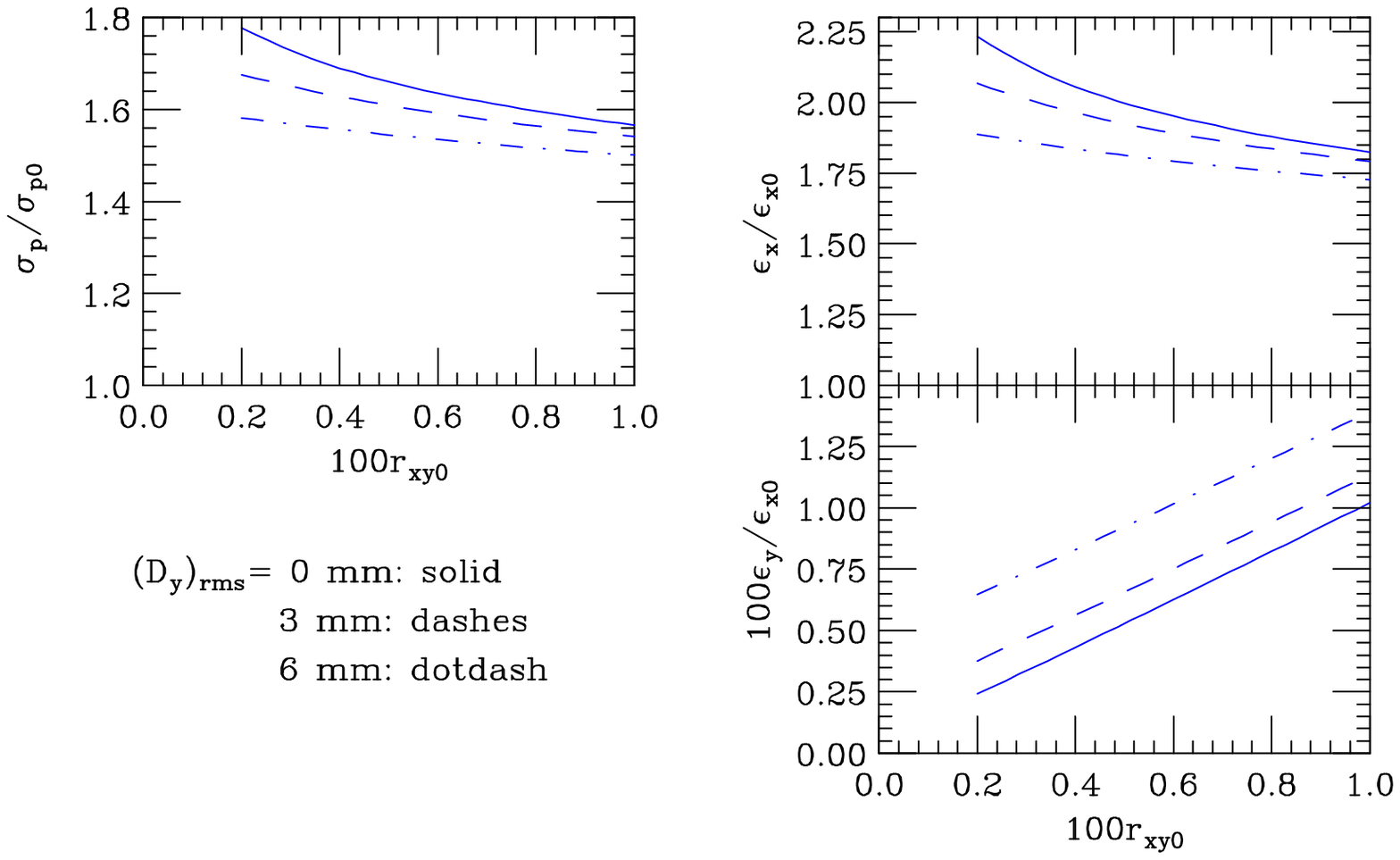, width=12.6cm}
\caption{
Calculations of the $r_{xy0}$ dependence of the beam
properties when $I=3.1$~mA, $f_{pw}=1$.
%Curves are given for random vertical dispersion
%with 3 different rms values, $(D_y)_{rms}=0$, 3, and
%6~mm.
}
\label{fidycoupv}
\end{figure}

\begin{figure}[p]
\centering
\epsfig{file=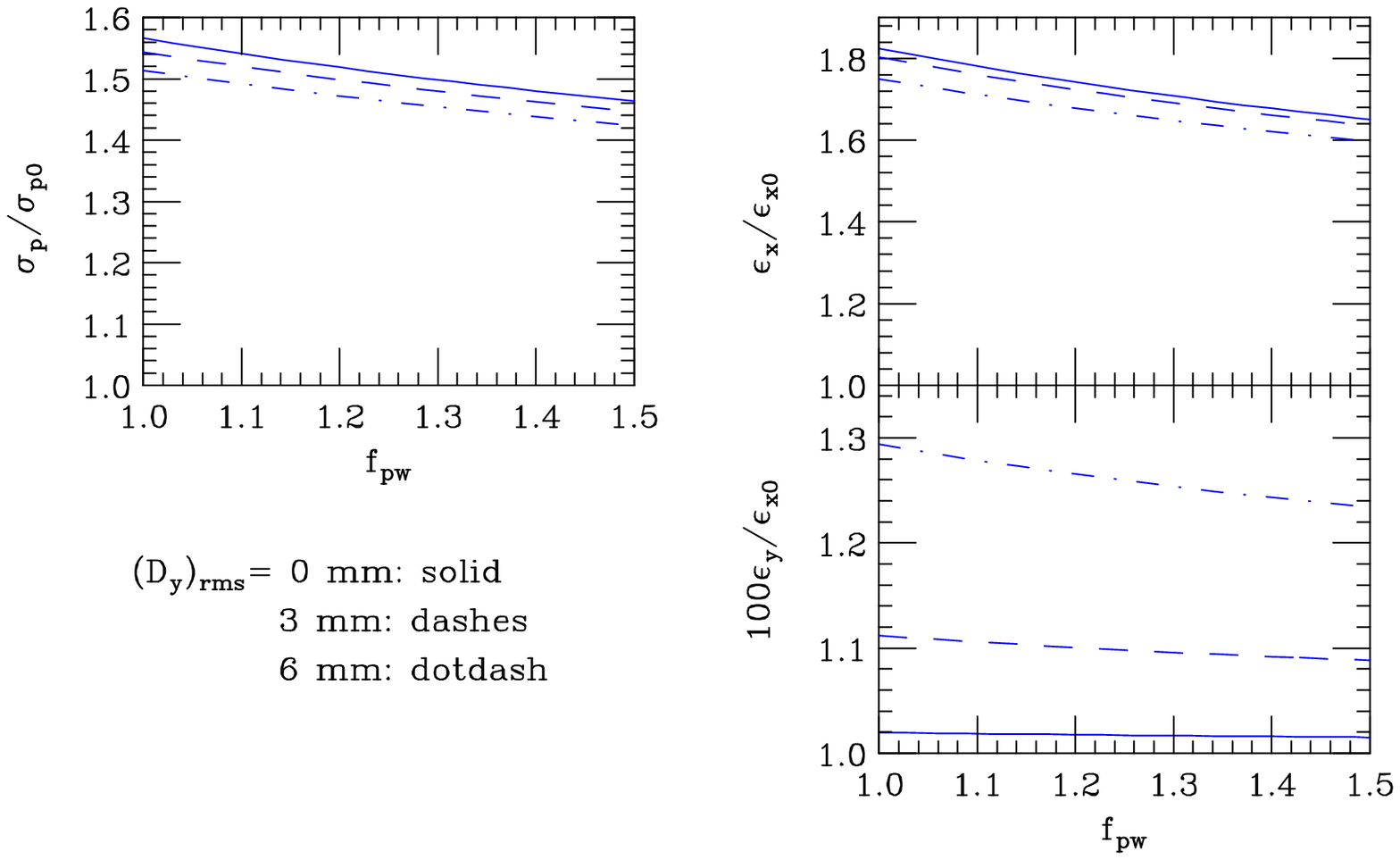, width=12.6cm}
\caption{
Calculations of the $f_{pw}$ dependence of the beam
properties when $I=3.1$~mA, $r_{xy0}=.01$.
}
\label{fifpotv}
\end{figure}

We see that the Piwinski and the Bjorken-Mtingwa methods give
essentially the same solution for the ATF, 
and since the P method allows for vertical dispersion,
we will choose
to continue our simulations using this method.
In Figs.~\ref{fidnv}-\ref{fifpotv} we give the steady-state 
emittance and energy spread {\it vs.} various parameters
in the ATF according to Piwinski's
IBS theory. We include curves representing $(D_y)_{rms}=0$, 3, and
6~mm. [Note that $(D_y)_{rms}=0$ is not a realistic condition for the ATF.]
Given the approximate nature of IBS theory, these results are meant
to give an idea of the sensitivities of the steady-state beam properties to
various parameters, and not to give absolutely correct predictions.
In Fig.~\ref{fidnv} we show the dependence on $I$, with
$f_{pw}=1$, $r_{xy0}=.01$. 
We note that the vertical emittance growth is almost zero with
zero vertical dispersion. At $(D_y)_{rms}=6$~mm, $\epsilon_y$
has increased by 44\% by $I=3$~mA.
In Fig.~\ref{fidycoupv} we show the dependence
on $r_{xy0}$, with $f_{pw}=1$, $I=3.1$~mA.
We note that at $r_{xy0}=0$ the vertical emittance does
not go to zero when $(D_y)_{rms}$ is not zero.
Note also that for the case $(D_y)_{rms}=6$~mm, the $\sigma_p(r_{xy0})$
curve is rather linear over the range shown, with a slope 
$\Delta\sigma_p/\sigma_p/\Delta r_{xy0}=6.6$; {\it i.e.} 
an (absolute) change in $r_{xy0}$ of 0.005 produces a relative change
in $\sigma_p$ of only 3\%. 
In Fig.~\ref{fifpotv} we give
the dependence on $f_{pw}$, with $I=3.1$~mA, $r_{xy0}=.01$.
Note that at $I=3$~mA, the ATF measurements give $f_{pw}=1.25$.
We see that the energy spread and emittances are not very
sensitive to this parameter. For example, a change of .1 in $f_{pw}$
yields a 1\% change in $\sigma_p$.
 
\begin{figure}[p]
\centering
\epsfig{file=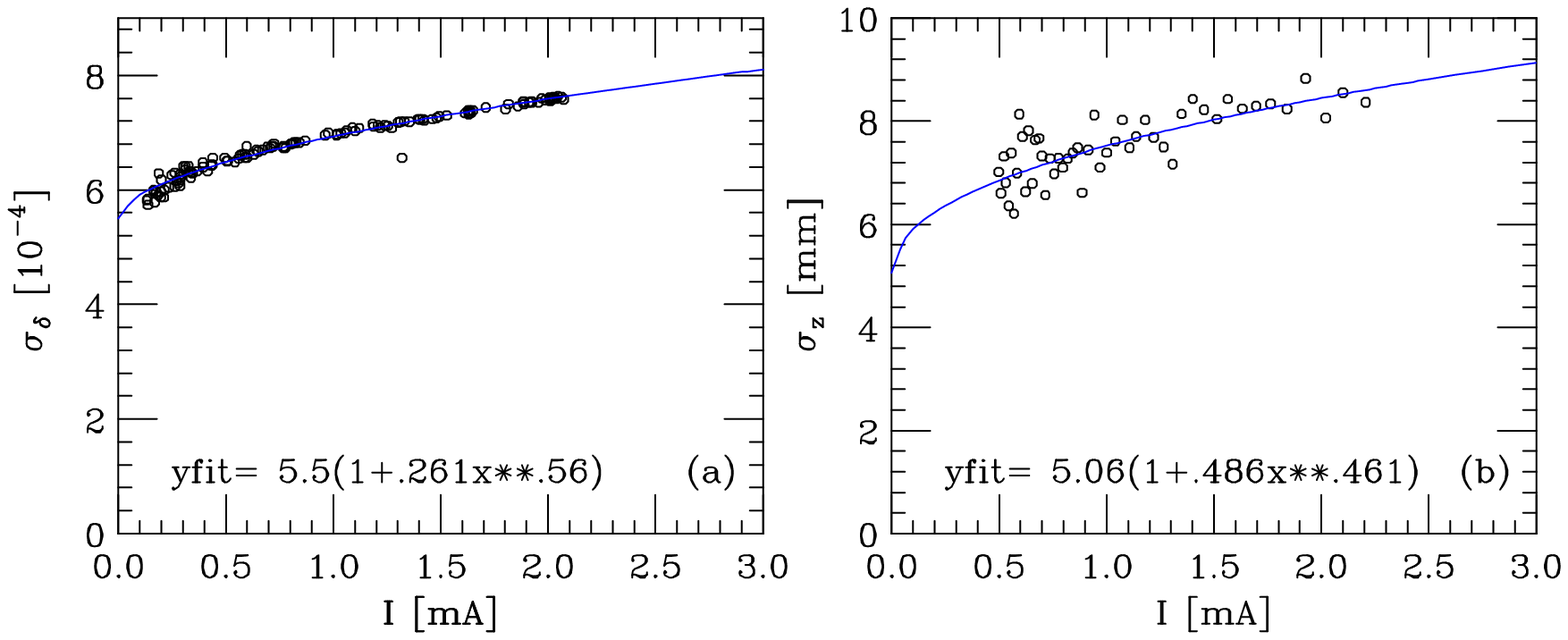, width=12.6cm}
\caption{
Measurements of energy spread~(a) and bunch length~(b),
 with $V_c=300$~kV, performed on April 13-14, 2000.
The fits (the curves) are chosen to give the expected zero
current results.
}
\label{fisigpsigz}
\end{figure}

\section*{Measurements\cite{april}-\cite{emittanceb}}

In April 2000, $\sigma_p$, $\sigma_s$, $\epsilon_x$, and $\epsilon_y$
were all measured as functions of current to varying degrees of accuracy. 
Among the most accurate is 
believed to be the energy spread measurement, which is performed 
on a thin screen at a high dispersion region 
in the extraction line ($D_x=1.73$~m). 
At different currents the measured beam width was fit to a Gaussian
(the fits were very good) and the rms width was extracted.
This measurement was performed on April 14 (see Fig.~\ref{fisigpsigz}a).
The rms scatter in
the extracted $\sigma_p$ was less than 2\%, and the precision should be better
than 1\%.
Note that
although data was obtained for currents up to $\sim2$~mA only,
from experience we have confidence that we can extrapolate to $\sim3$~mA.

 The rms bunch length, using
a streak camera, was also measured about the same time,
and there was more scatter in the
 data(see Fig.~\ref{fisigpsigz}b). With streak camera
measurements, however,
there is always 
the question of whe\-ther space charge in the streak camera
itself could have added a systematic error to the results. Checks 
were made with light filters, so we don't think this is a problem,
but this still adds a slight uncertainty to the results.
Both results were fit to a smooth curve, chosen to give the
expected zero current results (see Fig.~\ref{fisigpsigz}).
Note that these results, if true, imply an extremely large bunch
length increase at low currents.

The emittances were measured 
on wire monitors in the extraction line on April 19.
The results are reproduced by the plotting symbols in Fig.~\ref{fifit}b-c.
We see that the $x$ emittance appears to grow by $\sim80\%$
by $I=3$~mA; the
$y$ emittance begins at about 1\% of the $x$ emittance, and then grows
by a factor of 3.
We believe that 
the $x$-emittance measured
 should be fairly accurate. The $y$-emittance,
however, since it is so small, could be corrupted by many factors,
such as dispersion in the extraction line or the wire monitors (roll
of the measurement wires,
for example,
has been checked and shown not to be significant).
Note that it appears that $r_{xy0}\sim.01$. We estimated above,
using Eq.~\ref{epsy0}, that if the vertical emittance is dominated by
vertical dispersion, then $r_{xy0}=.01$ implies that $(D_y)_{rms}=7.5$~mm,
which is significantly larger than the measured 3-4~mm. Unfortunately,
we do not know what
$D_y(s)$ was during the April measurements, and therefore cannot
make a more precise comparison with calculation.

Note
that all the measurements were not performed on the same day, and
since IBS depends on the status of the machine ({\it e.g.} on the
vertical closed orbit distortion), it is possible that
the longitudinal and transverse measurements correspond to 
slightly different machines as far as IBS is concerned.
For example, on April 13, the day before the $\sigma_p$
measurements of Fig.~\ref{fisigpsigz}, the energy spread was
also measured. Those results, 
when fitted and extrapolated to $I=3$~mA, gave a 2\% smaller 
rms value than given here. 
This adds
some uncertainty to our results.

\section*{Comparison with IBS Theory}

\begin{figure}[htb]
\centering
\epsfig{file=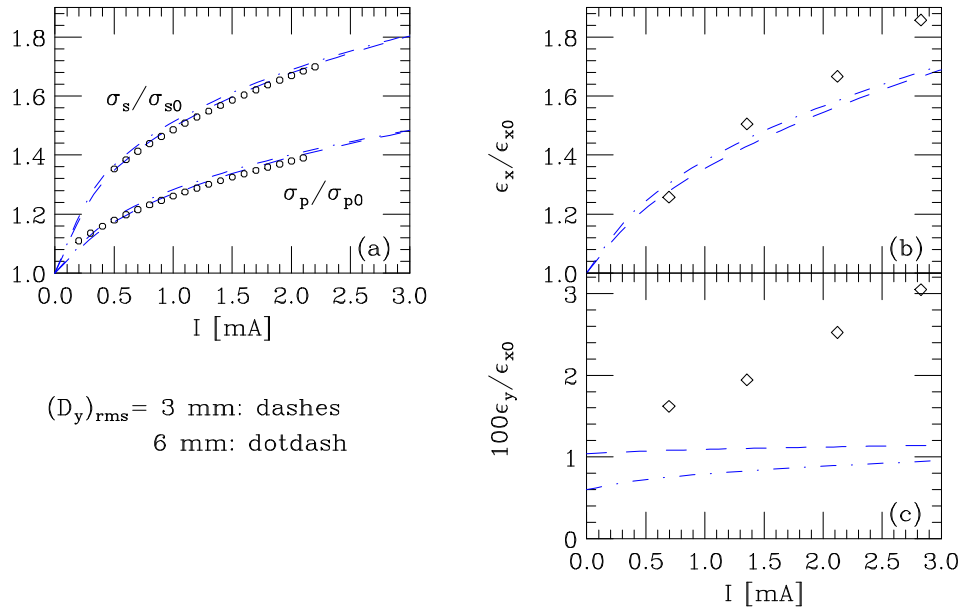, width=12.6cm}
\caption{
ATF measurement data (symbols) and IBS theory fits
(the curves). 
The symbols in (a) give the smooth curve fits
to the measured data of Fig.~\ref{fisigpsigz}.  
In the theory the parameter $r_{xy0}$ was adjusted 
to get the measured $\sigma_p$ at $I=3$~mA.
}
\label{fifit}
\end{figure}

To study the consistency of these measurements with IBS theory,
we perform
IBS calculations where we take
the potential well factor $f_{pw}$ that was measured, and adjust $r_{xy0}$  
to obtain 
the measured $\sigma_p$ (and $\sigma_s$) at $I=3$~mA, for the cases  
$(D_y)_{rms}=3$ and 6~mm.
[Remember: the typical measured value is $(D_y)_{rms}=3$-4~mm.]
The best fits were found for $r_{xy0}=.0104$ and .006, for the
cases $(D_y)_{rms}=3$ and 6~mm, respectively.
The results are shown in Fig.~\ref{fifit}, where they are compared
to the measured data.
(Note that in Fig.~\ref{fifit}a, we reproduce, using plotting symbols, the 
smooth curve fits
to the measured data of Fig.~\ref{fisigpsigz}.)

We first notice that
whether there is a 3~mm or 6~mm $y$ dispersion
the fitted results are very similar.
We find for our fitted results that the $\epsilon_x$ dependence on current
is in reasonable agreement with measurement (though 
the calculated result is low by 10\% at 3~mA), and
that $r_{xy0}\sim.006$-.01, which is a little low compared to .01.
The biggest discrepancy, however, seems to be that the vertical
emittance growth is much lower for the calculations than the measurements:
for the case $(D_y)_{rms}=3$~mm we see 10\% growth by 3~mA; for
$(D_y)_{rms}=6$~mm we see 60\% growth by 3~mA,
 which, however, is not close to the factor of 3 that was measured.
%We conclude, therefore that, if the measured vertical emittance
%growth is real, there must be some other effect, in addition
%to IBS, that contributes.
%Note also that when using the B-M method with $(D_y)_{rms}=0$,
%at $I=3$~mA, $\sigma_p$ increases by 1\%, $\epsilon_x$ increases
%by 4.5\%.
%Finally, note also from Fig.~\ref{fifit}
%that, other than the small current dependence of $\epsilon_y$,
%whether there is a 6~mm $y$ dispersion or no $y$ dispersion,
%the fitted results are very similar.
%One could reduce $r_{xy0}$ to get better agreement with
%the $\epsilon_x$ behavior, but then the resulting
%$\sigma_p$ and $\sigma_s$ would become too large.
% with this method, at 3.1~mA, 
%to get agreement with 
%the measured $\sigma_p$ and $\sigma_s$ requires that
%$c_{xy0}=6\%$ and $\epsilon_x/\epsilon_{x0}=1.36$, clearly wrong. 

Let us suppose that the 
measured vertical emittance growth is not real, 
and is due to measurement error.
In such a case,
our results appear to give reasonable agreement between IBS
theory and the ATF measurements, for $\sigma_p$, $\sigma_s$, and
$\epsilon_x$ {\it vs.} $I$,
 and low current $\epsilon_y$, all without extra 
fudge factors. 
%This suggests that, maybe, 
%the vertical emittance growth is not real, and is just due to 
%measurement error.
%If this were indeed the case, then
%This suggests that for the ATF, IBS theory might be a useful
%beam diagnostic. For example, from a measurement of $\sigma_p$,
%a relatively easy measurement to perform, one might be able to find
%$r_{xy0}$, which is a more difficult quantity to obtain directly. 
For the present set of data, however, we are still left with
some uncertainties.
For example, it was suggested earlier that the accurate knowledge of
$\sigma_p(I)$ is important. It was shown that for 
a realistic type of vertical dispersion for the ATF, $(D_y)_{rms}=6$~mm,
at $I=3.1$~mA,  
a change in $r_{xy0}$ of .005 produces a relative change
in $\sigma_p$ of only 3\%. Or, conversely, if the inaccuracy
in $\sigma_p(3~mA)$ is 3\%, then the fitted $r_{xy0}$ shifts by .005,
which is not small compared to .01.
And from the measured difference in $\sigma_p$ on April 13 and
14, it seems that the uncertainty in $\sigma_p$ is of this order.
Finally, we should
note that another source of uncertainty is that we don't know 
the function $D_y(s)$ on
the date of the measurements.

%Two voltages: expected difference= 4\%, calculated 1\%. 
%Tor's results.

However, there is one major quantative discrepancy between theory and
measurement, which has to do with
non-Gaussian beam tails. 
SAD finds that for the ATF
the IBS induced emittance growth, when not counting such tail particles,
is only 2/3 of that when all particles are included
(or 80\% in the case of $\sigma_p$), in agreement with
calculations for the ATF given in Ref.~\cite{Tor}.
Our calculations include all beam particles, while the measurements
consider only core particles.
If the tail particles are indeed as significant
as these results suggest, then 
the effect measured at the ATF is much stronger than predicted
by IBS theory, and there would no longer be good
agreement with measurement.
If one wants to think about a fudge factor that is multiplied
with the bunch charge, then we estimate that this factor must
be $\sim1.5$-2.0 (though no single such factor gives
good agreement for all $\sigma_p(I)$, $\sigma_s(I)$, and $\epsilon_x(I)$).
From this result, it appears that
either there is another, unknown force in addition to IBS causing
emittance growth in the ATF,
or the problem is the approximate nature of IBS theory.

In spite of this discrepancy, we believe that the IBS effect
will be useful as a diagnostic at the ATF. 
%Before taking any quantitive IBS results
% too seriously, however, we should remember
%that IBS theory is only approximately correct.
However, before we can use it as such, we need to
perform more
benchmarking experiments.
Such measurements include the effects of varying the rf voltage,
the vertical dispersion, and the
$x$-$y$ coupling. Additional independent measurements are also
desireable, such as using the interferometer to measure the
beam sizes in the ring. The calculations can also be improved by
including a cut-off for the tails.
In addition, one might, for example,
include the measured vertical dispersion (instead of a randomly
generated one) and/or use Piwinski's more involved formulation
that properly includes linear coupling. 

\section*{Conclusion}

Our theoretical comparisons
suggest that the ATF measurement results of April 2000 for 
energy spread, bunch length, and horizontal emittance {\it vs.}
current, 
 and a low current emittance ratio of about
1\% are generally consistent with intrabeam scattering (IBS) theory,
though the measured effects appear to be 
stronger than theory. In particular, 
the measured factor of 3
growth in vertical emittance at 3~mA
does not seem to be supported.
It appears that 
either (1)~there is another, unknown force in addition to IBS causing
emittance growth in the ATF;
or (2)~the factor of 3 vertical emittance growth is not real, and
our other discrepancies
are due to the approximate nature of IBS theory.
Our results suggest, in addition, that, even if IBS theory
has inaccuracies, the effect will be
useful as a diagnostic in the ATF.
For example, by measuring the energy spread, one
will be able to obtain the emittances.
Before this can be done, however, 
more benchmarking measurements will be needed.

\section*{Acknowledgements}

We thank A. Piwinksi for many useful comments and explanations
about the IBS effect. We thank Y. Nosochkov for lattice help,
and C. Steier for supplying the ALS lattice to compare with.
One of the authors (K.B.) thanks the ATF scientists and staff
for their hospitality and help during his visits to the ATF.

\end{document}